\documentclass[twocolumn,prd,showpacs,preprintnumbers,amsmath,amssymb]{revtex4}
\usepackage{graphicx}

\begin{document}

\newcommand{\fig}[1]{Fig.~\ref{#1}}
\newcommand{\eq}[1]{Eq.~\ref{#1}}

\title{Study of an electro-optic modulator capable of generating
        simultaneous amplitude and phase modulations}

\author{Benedict J Cusack, Benjamin S Sheard, Daniel A Shaddock
  \footnote{Daniel Shaddock is currently employed at the Jet
Propulsion Laboratory, Pasadena CA, USA.}, Malcolm~B~Gray and Ping
Koy Lam} \affiliation{Department of Physics and Theoretical
Physics, Faculty of Science,
\\The Australian National University, Canberra ACT 0200, Australia}
\author{Stan E Whitcomb}
\affiliation{LIGO Laboratory, MS 18-34, California Institute of
Technology, Pasadena CA, USA}

\date{Friday, 8th August, 2003}

\begin{abstract}
We report on the analysis and prototype-characterization of a
dual-electrode electro-optic modulator that can generate both
amplitude and phase modulations with a selectable relative phase,
termed a \emph{universally tunable modulator} (UTM). All
modulation states can be reached by tuning only the electrical
inputs, facilitating real-time tuning, and the device is shown to
have good suppression and stability properties. A mathematical
analysis is presented, including the development of a geometric
phase representation for modulation. The experimental
characterization of the device shows that relative suppressions of
38~dB, 39~dB and 30~dB for phase, single-sideband and
carrier-suppressed modulations, respectively, can be obtained, as
well as showing the device is well-behaved when scanning
continuously through the parameter space of modulations.  Uses for
the device are discussed, including the tuning of lock points in
optical locking schemes, single sideband applications, modulation
fast-switching applications, and applications requiring combined
modulations.

\end{abstract}

\maketitle

\section{Introduction}

Although electro-optic amplitude modulators and phase modulators
are commonplace in modern optics laboratories, there is no single,
commercially available device that produces controllable amplitude
and phase modulations with complete variability.  When required,
various optical configurations have been assembled, such as phase
and amplitude modulators in series \cite{pmam}, and two phase
modulators used in a Mach-Zehnder interferometric setup (the
Mach-Zehnder modulator) \cite{mzm}, which are both capable of
amplitude and phase modulation.

The device discussed in this paper, called a \emph{universally
tunable modulator} (UTM), is a modification of a commercial
electro-optic amplitude modulator. An amplitude modulator consists
of two phase-modulation crystals aligned at right angles, and
connected with opposite polarities to a single electrical input. A
UTM (see \fig{figsmvectors}(a)) is identical, but with two
separate inputs, one connected to each crystal.  A prototype was
constructed from a New Focus broadband amplitude modulator (Model
$4104$) in this way.  We demonstrate here that this extra degree
of freedom enables the production of phase as well as amplitude
modulation.

We use the term "universally tunable" to highlight the fact that
the device is capable of both amplitude modulation (AM) and phase
modulation (PM), with full selectability of the amplitude of each
and of the relative phase between the two. Choosing a figure of
merit with which to assess the device is difficult in general, and
depends on the modulation requirements of the application in
question.  We address this issue here by identifying two
characteristics that might be expected of the modulations states
produced by the device: \emph{variability} and \emph{purity}.  The
variable nature of the device is brought out by experiments which
require real-time tuning of the modulator across its parameter
space. Other experiments rely on precise selection of a small
number of operating points, where the purity of the modulations
produced has a direct influence on the success of the experiment.

One anticipated use for the UTM device is in optical feedback
control of the resonance condition of a Fabry-Perot cavity, or the
fringe condition of a Michelson interferometer. The RF modulation
techniques used to lock these devices are Pound-Drever-Hall
locking \cite{pdh} and Schnupp modulation locking \cite{schnupp},
respectively, and are both based on properties of the devices that
convert (injected) PM to AM, which is then demodulated to give an
error signal readout.  In both cases, the default locking point is
at a turning point in the transmitted (or reflected) optical field
intensity; injection into the device of additional AM (along with
the default PM) causes the device to lock with an offset relative
to the turning point. Moreover, the quadrature of AM required for
Pound-Drever-Hall locking is orthogonal to that required for
Schnupp modulation locking, introducing the possibility of
independently tuning two lock points at once in a coupled system.
This \emph{RF offset locking} application is our case study
example to demonstrate the variability property of a prototype
UTM; the technique may be useful for future large-scale
gravitational wave detector configurations \cite{advligo}, with
the feature of facilitating real-time tuning of detector frequency
responses \cite{sigrec,fpvrsm,mivrsm}.

The UTM device is capable of several "pure" states, which have
various applications as well as providing a natural way to test
the purity property of the prototype. Single-sideband modulation,
for example, requires an equal combination of AM and PM, in
quadrature.  The subject has arisen a number of times with optic
fibre technology where, for example, a single-sideband-modulated
signal is immune to fibre dispersion penalties \cite{mzm}, and the
technique has also been suggested for subcarrier-multiplexing
systems \cite{mplex}. Single-sideband modulation has been achieved
by other means: optically filtering out one sideband
\cite{opfilt}; cascaded amplitude and phase modulators
\cite{pmam}; Mach-Zehnder modulators \cite{mzm}; and more complex
arrangements \cite{sagnacm}. While these methods have merit, we
submit that the UTM is far simpler in design, with fewer degrees
of freedom available to drift, and is at least comparable
regarding suppression capabilities.

The UTM can produce PM or AM states, with purity limited by the
accuracy of polarisation optics (including the birefringent
effects of the device itself). While these states are obtainable
using off-the-shelf amplitude or phase modulators, applications in
coherent state quantum cryptography require fast-switching between
AM and PM \cite{switch}, for which the UTM is ideal.  Other
applications, including some quantum communication protocols
\cite{cwtel}, would also benefit from having easy access to a
tuned, stationary combination of AM and PM.  In addition, we show
that the UTM (or indeed an amplitude modulator) can produce a
\emph{carrier suppression} state, where the output consists only
of two modulation sidebands.

We have developed a geometric phase sphere description of
modulation states, the \emph{modulation sphere} formulation, in
analogy with the Poincar\'{e} sphere description for polarisation
states \cite{poincare}. We demonstrate the use of this formalism
by calculating the transfer function (electrical to optical) of
the UTM device, both using optical-field-phasors and using
modulation sphere parameters, and discuss the pros and cons of the
two mathematical descriptions.  We have found that the modulation
sphere representation is particularly useful for gaining a visual
insight into the dynamics of the UTM.

Section \ref{2model} gives the mathematical description of the
UTM, in terms of optical-field-phasors (Section \ref{2phasor2}),
and the modulation sphere formulation (Section \ref{2modsphere3}).
The transfer functions from these two subsections are derived in
Appendices \ref{secapp1} and \ref{secapp2} respectively.  Section
\ref{3expt} describes a characterisation experiment conducted on a
prototype UTM, with the experimental layout described in Section
\ref{3expt2}, and the results presented in Section
\ref{3results3}. The main results of the study are summarised in
Section \ref{4conc}.

\section{Theoretical Model}
\label{2model}

\subsection{Overview}
\label{2over1}

Before we proceed, a simple analogy may help to clarify the inner
workings of a UTM device, and lend physical insight.  A
Mach-Zehnder modulator involves splitting a beam into two parts,
separately phase-modulating each part, then recombining the two
parts on a beamsplitter.  The UTM is physically equivalent to
this, where the two interferometer paths are collinear but
different in polarisation.  Each UTM crystal modulates one and
only one polarisation component.  The phase between the electric
fields in the two polarisation states we identify as the
interferometric recombination phase.  The birefringent waveplates
used to create the initial polarisation take the place of the
input beamsplitter, and a polarising beamsplitter facilitates the
output recombination. The recombination phase is adjustable (by
modifying the polarisation state), and one can control the
amplitudes and phases of the single-beam phase-modulations
generated by each of the two crystals.  Through manipulation of
these input parameters, the user has complete control over the
interference condition of the two carrier beams, and separately
over the interference of the lower and upper sidebands generated
by the modulating crystals. This amounts to having complete
freedom to choose any modulation state by appropriate choice of
the input parameters.

As an example, consider two identically phase-modulated beams,
where the optical phase of one beam is advanced by $90^{\circ}$,
and the modulation phase of the other beam is advanced by
$90^{\circ}$. The result, upon interfering the two beams, is that
one resultant sideband is exactly cancelled out and the other is
additively reinforced, producing a single sideband state.  One can
intuitively arrive at all of the modulation states discussed in
this paper by reasoning along these lines.  In the next section,
we quantify such reasoning into a mathematical theory of the UTM.

\subsection{UTM Transfer Function}
\label{2phasor2}

\begin{figure}[!t]
  \centerline{\includegraphics{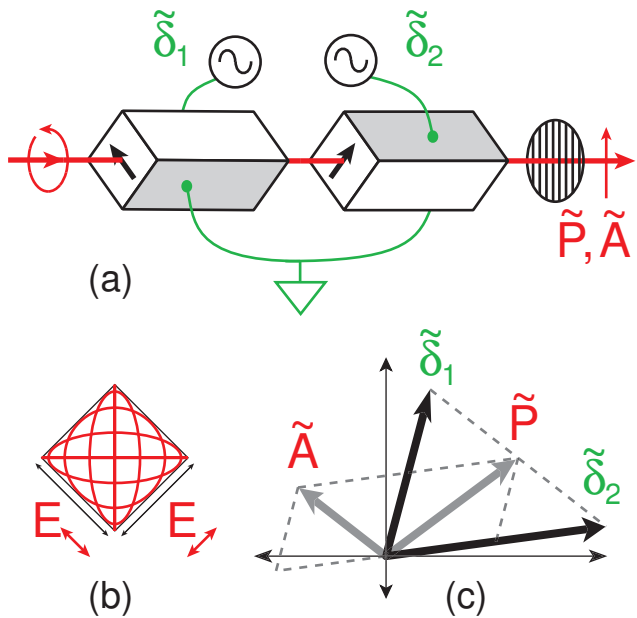}}
  \caption{\footnotesize{
        (a) UTM schematic.  Two modulating crystals are positioned in
        series with modulating axes at $90^{o}$.  A laser beam of
        elliptical polarisation passes through each crystal and
        then through a vertically polarising element.
        (b) The subset of input beam polarisation states
        considered here is all states such that there is an equal optical field
        amplitude in both the left- and right-diagonal components.
         $\sigma$ equals the phase between these components: shown
         are $\sigma = 0$ (vertical), $\pi/4$, $\pi/2$
         (circular), $3\pi/4$, and $\pi$ (horizontal).
        (c) Phasor diagram representing the transfer function
        given by \eq{eqntf}. If the input electrical signals are
        represented as rotating phasors ($\tilde{\delta}_1$ and
        $\tilde{\delta}_2$), then the corresponding phasors that
        represent the output phase and amplitude modulations
        ($\tilde{P}$ and $\tilde{A}$) are proportional to their sum
        and difference respectively.}}
    \label{figsmvectors}
  \end{figure}

As stated earlier, a UTM (see \fig{figsmvectors}(a)) consists of
two modulating crystals in series, with separate voltage sources,
and with modulating axes at right angles to each other. The UTM is
used by sending a laser beam of elliptical polarisation through
the system, and then through a linearly polarising filter angled
at $45^{\circ}$ to the crystals' modulating axes.

The choice of polarisation state of the incident beam has a
crucial effect on the transfer characteristics of the system.  We
will restrict ourselves to a subset of polarisation states: those
where there is equal power in the polarisation components aligned
with each modulating crystal axis.  Equivalently, we require that
the polarisation is elliptical with major and minor axes aligned
at $45^{\circ}$ to both crystal axes.  For concreteness we will
choose the ellipse axes to be horizontal and vertical, and the
crystal axes to be left and right diagonal (as in
\fig{figsmvectors}(a)). We define the angle $\sigma$ as being the
phase difference between left-diagonal and right-diagonal electric
field components of the light exiting the crystals (see
\fig{figsmvectors}(b)). The identity
\begin{equation}
  \label{eqnident}
  \mathrm{tan}^2(\frac{\sigma}{2}) = \frac{P_h}{P_v}
\end{equation}
follows, where $P_{h,v}$ are the powers in the horizontal and
vertical polarisation components of the input laser beam.  We
introduce this degree of freedom with the application of
gravitational wave detector locking schemes in mind, where laser
power is at a premium: choosing $\sigma$ such that the beam is
almost vertically polarised allows us to retain the majority of
the input carrier power, while retaining full access to all
modulation states, albeit with reduced amplitude.

Note that, in the ideal situation where both crystals have
identical refractive indices and lengths, the phase angle $\sigma$
(and hence the polarisation state) is the same before and after
the modulator.  However, a real device almost certainly does not
have identical crystals so that the relative phase between left-
and right-diagonal components may well be changed by the crystals,
in which case $\sigma$ should describe the polarisation state of
the light \emph{exiting} the modulator, just before it reaches the
linearly polarising filter.

The inputs to the two crystal electrodes are sinusoidal voltages
of frequency $\omega_m$; we describe these as complex phasors
$\tilde{\delta}_j = \delta_j e^{i\phi_j}$ ($j = 1,2$, and
$i=\sqrt{-1}$) so that the single-crystal phase modulations are
given by $\Re \{\tilde{\delta}_j e^{i\omega_m t}\} = \delta_j
\mathrm{cos} (\omega_m t + \phi_j)$, where $\Re$ and $\Im$
respectively return the real and imaginary components of a complex
variable.

We can similarly write the output phase and amplitude modulations
as complex phasors $\tilde{P}~=~P~e^{i\phi_P}$ and
$\tilde{A}~=~A~e^{i\phi_A}$ where the optical field amplitude
exiting the UTM can be shown to be:
\begin{eqnarray}
  \label{eqnelecfield}
  E_{\mathrm{OUT}} = E_{\mathrm{IN}} \, \left[ \,
  \mathrm{cos}(\sigma/2) + i \Re \{ \tilde{P} e^{i\omega_mt}\}
  + \Re \{ \tilde{A} e^{i\omega_mt}\} \right]  \nonumber \\
  = E_{\mathrm{IN}} \, \left[ \, \mathrm{cos}(\sigma/2)
    + i P \mathrm{cos}(\omega_m t + \phi_P) + A \mathrm{cos}(\omega_m t + \phi_A) \, \right]
\end{eqnarray}
where $E_{\mathrm{IN}}$ is the input optical field amplitude.  In
\eq{eqnelecfield}, we have assumed small single-crystal modulation
depths ($\delta_j \ll 1$), and we have factored away the net
optical phase dependence.

The relationship between input and output parameters for the UTM
device can be shown to be:
\begin{eqnarray}
  \label{eqntf}
  \tilde{P} &=& \frac{1}{2} \, \mathrm{cos}\left(\frac{\sigma}{2}\right)\left( \tilde{\delta}_1 +
  \tilde{\delta}_2 \right) \nonumber \\
  \tilde{A} &=& \frac{1}{2} \, \mathrm{sin}\left(\frac{\sigma}{2}\right)\left( \tilde{\delta}_1 -
  \tilde{\delta}_2 \right).
\end{eqnarray}
(A proof of this transfer function is outlined in Appendix
\ref{secapp1}.)

An important observation from \eq{eqntf} is that the phases of the
AM and PM track the phases of the sum and difference of the
electrical signals, respectively. Thus, an electronic oscillator
for use in demodulation schemes is readily available, and can be
calibrated once for all operating points. This phase-tracking
property holds for any choice of $\sigma$; the effect of a change
of polarisation ellipticity is to change the overall quantity of
PM or AM available (as well as changing the overall beam power in
\eq{eqnelecfield}).

Returning briefly to the topic of conserving as much carrier power
as possible for gravitational wave detector applications, we
derive a useful rule-of-thumb for selecting an appropriate amount
of power to tap off.  Consider \eq{eqntf} when we set
$\tilde{\delta}_1 = i\tilde{\delta}_2$ such that $\tilde{\delta}_1
+ \tilde{\delta}_2$ and $\tilde{\delta}_1 - \tilde{\delta}_2$ have
the same magnitude (we do this in order to focus on the
\emph{coefficients} of the complex $\tilde{\delta}_j$ phasors).
The ratio of modulation power between the AM and PM components is
$|\tilde{A}|^2/|\tilde{P}|^2 = \mathrm{tan}^2(\sigma /2)$ and,
comparing this with \eq{eqnident}, we have:
\begin{equation}
   \label{eqnthumb}
   \frac{|\tilde{A}|^2}{|\tilde{P}|^2} = \frac{P_h}{P_v}
\end{equation}
So, the fraction of power tapped off is directly proportional to
the fraction of modulation power available to be expressed as AM
(rather than PM). Typically then, if $10\%$ of the output power is
tapped off, and the user wishes to alternately produce a pure PM
state and then a pure AM state (each of the same modulation
power), then the AM state will require 9 times the input
electrical power in compensation, compared to the PM state.

It is apparent from Eqs. \ref{eqnelecfield} and \ref{eqntf} that
for $\sigma = \pi$ the carrier and the PM contribution vanish,
leaving only the AM component.  This is the \emph{carrier
suppression} operating point, where only two sidebands remain. The
modulation ceases to be of the phase or amplitude type due to the
absence of a carrier as a phase reference, though the modulation
retains its beat phase through the second harmonic in optical
power.

A rearrangement of \eq{eqnelecfield} serves to clarify the nature
of single sideband modulation:
\begin{eqnarray}
  \label{eqnelecfield2}
  E_{\mathrm{OUT}} = E_{\mathrm{IN}} \, \left[ \,
  \mathrm{cos}(\sigma/2) + \frac{1}{2} \left( \tilde{A} + i
  \tilde{P} \right) e^{i\omega_mt} \right. \nonumber \\
  + \left. \frac{1}{2} \left( \tilde{A} - i
  \tilde{P} \right)^* e^{-i\omega_mt} \right]
\end{eqnarray}
where the $e^{\pm i\omega_mt}$ terms represent optical sidebands
at $\pm \omega_m$ relative to the optical carrier frequency, and
an asterisk denotes a complex conjugate.  It is clear then that
choosing $\tilde{A} = \pm i\tilde{P}$ eliminates one or other of
the sidebands.  In other words, a single sideband state
corresponds to having equal amounts of PM and AM where the two
modulations are in quadrature.

\subsection{Modulation Sphere Formulation}
\label{2modsphere3}

We now define a geometric phase representation of the modulation
parameters of the system to further clarify the transfer
characteristics of the UTM device. This is analogous to using
Stokes parameters and the Poincar\'{e} sphere representation for
optical polarisation states \cite{poincare}.  In this case, we
transform from the complex quantities $\tilde{P}$ and $\tilde{A}$
to a set of (real) coordinates $(M_1,M_2,M_3)$ in \emph{M-space}
via the transformation:
\begin{eqnarray}
  \label{eqngeospherem}
  M_1 &=& P^2 - A^2 \nonumber \\
  M_2 &=& 2 P A \mathrm{cos} (\phi_P - \phi_A) = 2\Re \{ \tilde{P} \tilde{A}^* \} \nonumber \\
  M_3 &=& 2 P A \mathrm{sin} (\phi_P - \phi_A) = 2\Im \{ \tilde{P} \tilde{A}^* \}
\end{eqnarray}

A single point in M-space represents a distinct modulation state.
In particular, the M-space representation suppresses the common
(beat) phase of the modulation state; the equations are defined in
terms of the relative phase difference between AM and PM
contributions. This is a useful simplification in that now every
physically distinct modulation corresponds to one point and one
only.  On the other hand, one cannot use the representation for
demodulation-phase-tracking calculations, when the physical
significance of the modulation's overall phase is renewed by the
presence of an external (electrical) phase reference.

 \begin{figure}[!t]
   \centerline{\includegraphics{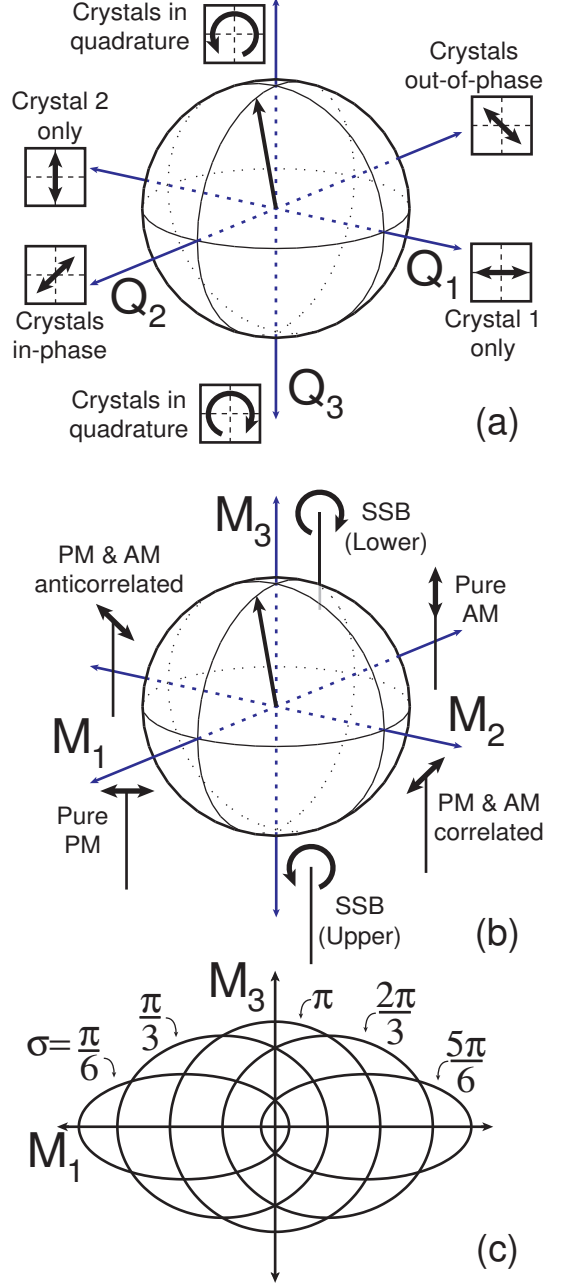}}
  \caption{\footnotesize{
        (a) Q-space diagram.  Axis labels are visual reminders
        of relative amplitudes and phases of input electrical
        signals (try to visualise the display of a cathode ray
        oscilloscope run in X-Y mode with $\delta_1$ and
        $\delta_2$ as the X and Y inputs).
        (b) M-space diagram.  Axis labels are optical amplitude
        phasors whose endpoints oscillate at the modulation
        frequency.  For $\sigma=\pi/2$, there is a graphical
        transfer correspondence between (a) and (b) according to
        \eq{eqntfsphere}. (c) In general, the transfer function of
        the UTM, \eq{eqntfellipse}, takes a sphere of constant
        input power in Q-space and returns an ellipse in M-space.
        A cross section is shown for a selection of values for
        $\sigma$.}}
    \label{figspheres}
  \end{figure}

\begin{widetext}
\begin{center}
\begin{table}
  \label{taboppoints}
  \begin{center}
  \scriptsize{
  \begin{tabular}{|c|c|c|c|c|}
    \hline
    Operating point & Modulations & Electrical Inputs & M-space &
    Q-space \\
    \hline
    Pure PM & $\tilde{P} \ne 0, \tilde{A} = 0$ & $\tilde{\delta}_1 = \tilde{\delta}_2$ &
    $(+M_0,0,0)$ & $(0,\frac{4M_0}{\mathrm{cos}(\sigma)+1},0)$ \\
    Pure AM & $\tilde{P} = 0, \tilde{A} \ne 0$ & $\tilde{\delta}_1 = \tilde{\delta}_2e^{i\pi}$ &
    $(-M_0,0,0)$ & $(0,\frac{4M_0}{\mathrm{cos}(\sigma)-1},0)$ \\
    Correlated PM \& AM & $\tilde{P} = \tilde{A}$ & $\tilde{\delta}_1 \mathrm{cos}(\sigma)
    = \tilde{\delta}_2 e^{i\pi} (1 + \mathrm{sin}(\sigma))$ & $(0,+M_0,0)$ &
    $(\frac{4M_0}{\mathrm{sin}(\sigma)},-\frac{4M_0\mathrm{cos}(\sigma)}{\mathrm{sin}^2(\sigma)},0)$ \\
    Anticorrelated PM \& AM & $\tilde{P} = \tilde{A}e^{i\pi}$ &
    $\tilde{\delta}_1 (1 + \mathrm{sin}(\sigma)) = \tilde{\delta}_2 e^{i\pi} \mathrm{cos}(\sigma)$ &
    $(0,-M_0,0)$ & $(-\frac{4M_0}{\mathrm{sin}(\sigma)},-\frac{4M_0\mathrm{cos}(\sigma)}{\mathrm{sin}^2(\sigma)},0)$\\
    Single sideband $-\omega_m$ & $\tilde{P} = \tilde{A}e^{i\pi/2}$ & $\tilde{\delta}_1
    = \tilde{\delta}_2 e^{i(\pi+\sigma)}$ & $(0,0,+M_0)$ & $(0,-\frac{4M_0\mathrm{cos}(\sigma)}{\mathrm{sin}^2(\sigma)}
    ,-\frac{4M_0}{\mathrm{sin}(\sigma)})$\\
    Single sideband $+\omega_m$ & $\tilde{P} = \tilde{A}e^{-i\pi/2}$ & $\tilde{\delta}_1
    = \tilde{\delta}_2 e^{i(\pi-\sigma)}$ & $(0,0,-M_0)$ & $(0,-\frac{4M_0\mathrm{cos}(\sigma)}{\mathrm{sin}^2(\sigma)}
    ,\frac{4M_0}{\mathrm{sin}(\sigma)})$ \\
    \hline
  \end{tabular}
  }
  \caption{Details of significant operating points.  The modulation
  parameters, expressed both in terms of phasors and
  M-space parameters, are given for each point.  The electrical
  inputs required to produce these operating points are also given
  (both in terms of phasors and Q-space parameters),
  and these inputs vary with the polarisation parameter $\sigma$.}
  \end{center}
\end{table}
\end{center}
\end{widetext}

The three M-space parameters can be interpreted as follows: $M_1$
measures the extent to which one kind of modulation (AM or PM)
dominates over the other; $M_2$ measures the degree to which the
present PM and AM components are correlated or anti-correlated in
phase; and $M_3$ measures the degree to which the present PM and
AM components are in quadrature phase to each other, and thus also
measures the extent to which one frequency sideband has more power
than the other.  We also define the \emph{modulation power} by
$M_0 = P^2 + A^2$, from whence $M_0^2 = M_1^2 + M_2^2 + M_3^2$
follows. Therefore, a surface of constant $M_0$ is a sphere in
M-space (see \fig{figspheres}(b)).

We attach particular significance to modulation states whose
M-space coordinates lie on one cardinal axis.  The point
$(+M_0,0,0)$ represents pure PM, and the point $(-M_0,0,0)$
represents pure AM.  The points $(0,\pm M_0,0)$ represent
correlated and anti-correlated PM and AM, and $(0,0,\pm M_0)$
represent upper and lower single sideband states.  Details of
these states are summarised in Table I.  These states,
particularly those on the $M_1$ and $M_3$ axes, involve the
suppression of a particular signal, and so provide a natural way
to test the precision of a real device.

Before we proceed, it is convenient to define an analogous
\emph{Q-space} representation, $(Q_1,Q_2,Q_3)$, for the electrical
input parameters, via:
\begin{eqnarray}
  \label{eqngeospherep}
  Q_1 &=& \delta_1^2 - \delta_2^2 \nonumber \\
  Q_2 &=& 2 \delta_1 \delta_2 \mathrm{cos} (\phi_1 - \phi_2) = 2\Re \{ \tilde{\delta}_1 \tilde{\delta}_2^* \} \nonumber \\
  Q_3 &=& 2 \delta_1 \delta_2 \mathrm{sin} (\phi_1 - \phi_2) = 2\Im \{ \tilde{\delta}_1 \tilde{\delta}_2^* \}
\end{eqnarray}
where $Q_0 = \delta_1^2 + \delta_2^2 = \sqrt{Q_1^2 + Q_2^2 +
Q_3^2}$ is the sum input electrical power. Physical
interpretations for these parameters correspond with the M-space
equivalents (see \fig{figspheres}(a) for a diagrammatic
description).

Now, we can rewrite the UTM transfer function, \eq{eqntf}, in
terms of the M-space and Q-space parameters:
\begin{eqnarray}
  \label{eqntfellipse}
  M_0 &=& \frac{1}{4} \left(Q_0 + Q_2 \mathrm{cos}(\sigma) \right) \nonumber \\
  M_1 &=& \frac{1}{4} \left(Q_0 \mathrm{cos}(\sigma) + Q_2 \right) \nonumber \\
  M_2 &=& \frac{1}{4} \mathrm{sin}(\sigma) Q_1 \nonumber \\
  M_3 &=& -\frac{1}{4} \mathrm{sin}(\sigma) Q_3
\end{eqnarray}
(A proof of this transfer function is outlined in Appendix
\ref{secapp2}.)

Since the modulation power $M_0$ depends partly on $Q_2$ (and
hence depends on the \emph{detail} of the input signals, not just
the overall input power $Q_0$), a sphere of constant electrical
input power in Q-space does \emph{not}, in general, map to a
sphere of constant modulation power in M-space.  In fact,
\eq{eqntfellipse} describes a transfer function from a sphere in
Q-space to an \emph{ellipsoid} in M-space.  It can be shown that
the ellipsoid is centered at $(M_1,M_2,M_3) = (Q_0
\mathrm{cos}(\sigma) /4,0,0)$, has radii $[Q_0/4, Q_0
\mathrm{sin}(\sigma)/4, Q_0 \mathrm{sin}(\sigma)/4]$, and has an
eccentricity of $\epsilon = |\mathrm{cos}(\sigma)|$.  The ellipse
is always prolate with long axis aligned with the $M_1$-axis, and
always has one focus at the M-space origin.  The orthogonality of
axes is preserved in the transformation, which consists only of a
translation (the ellipsoid is not origin-centered) and a dilation
(the ellipsoid is squashed in the $M_2$ and $M_3$ directions). The
centrepoint and proportions of the ellipsoid are parametrised by
the input light's polarisation parameter, $\sigma$ (see
\fig{figspheres}(c)).

A special case occurs for $\sigma=\pi/2$ (circular input
polarisation), when \eq{eqntfellipse} reduces to:
\begin{eqnarray}
  \label{eqntfsphere}
  M_0 &=& \frac{1}{4} Q_0 \nonumber \\
  M_1 &=& \frac{1}{4} Q_2 \nonumber \\
  M_2 &=& \frac{1}{4} Q_1 \nonumber \\
  M_3 &=& -\frac{1}{4} Q_3
\end{eqnarray}

Here, a sphere in Q-space maps to a sphere in M-space.  This case
is particularly instructive as Figs. \ref{figspheres}(a) and
\ref{figspheres}(b) take on a new significance: there is now a
direct graphical correspondence between the two spheres, in
accordance with \eq{eqntfsphere}.  So, as an example of using
these spheres as a visual tool, we can see "at a glance" that the
PM operating point is obtained by running both crystals with equal
in-phase electrical inputs, or that a single sideband is obtained
by running both crystals with equal inputs in quadrature.


\section{Experimental Demonstration}
\label{3expt}

\subsection{Approach}

The following characterisation experiment was designed to measure
the amplitude and phase response of a prototype UTM device with
respect to both the AM and PM output states.  In particular, we
focus here on two kinds of measurement: those that measure the
\emph{variability} of the UTM device, and those that measure the
\emph{purity} of the modulations that the UTM device can produce.
By variability, we refer to the capability of the UTM prototype to
tune around the parameter space of modulations, or "dial up" a
particular modulation, in a predictable, theory-matching fashion
and with a reasonable level of precision.  By purity, we speak of
the UTM prototype's ability to accurately attain a particular
modulation (especially those that involve suppression of a
frequency line) and to hold that modulation indefinitely without
drifting.

\subsection{Experimental Layout}
\label{3expt2}

\begin{figure}[!t]
   \centerline{\includegraphics{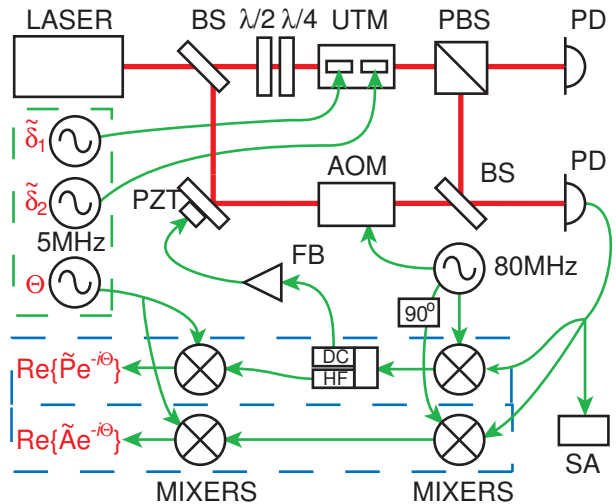}}
   \caption{\footnotesize{
        UTM heterodyne characterisation experiment layout.  A spectrum analyser
        provides frequency line data (of the direct AM beat, and
        heterodyned copies of sidebands) for establishing the purity of the UTM
        , while two parallel double-demodulation circuits provide
        the complete set of modulation measurables for testing
        the variability of the UTM.  BS: dielectric
        beam splitter; $\lambda/2$ half-wave plate; $\lambda/4$
        quarter-wave plate; PBS: polarising beam splitter; PZT:
        piezo-electric actuator; AOM: acousto-optic modulator; FB:
        feedback servo and HV amp; $90^{\circ}$: electronic
        phase-shifter; DC/HF: bias-T; SA: spectrum analyser. The
        5~MHz signal generators are electronically phase-locked.
        }}
    \label{figexpt}
  \end{figure}

The experiment is shown in \fig{figexpt}.  The polarisation of a
laser source was prepared with a half- and quarter-wave plate in
series, and the laser source was passed through the UTM, which we
operated at $\omega_{m}/2\pi = 5$~MHz. A polarising beam splitter
completed the process; the vertically polarized output carried the
modulation described in the theory above.

The horizontally polarised output was used to keep track of the
polarisation state, and hence to measure the value of $\sigma$.
The polarisation state used corresponded to $\sigma = 75^{\circ}$
(near-circular polarisation with the vertical component stronger),
for experimental convenience.

Regarding the modulator itself, the original New Focus amplitude
modulator (Model 4104) carried the specification: max $V_\pi =
300$~V @ $1.06\:\mu\mathrm{m}$.  While we did not explicitly
measure $V_\pi$ for the prototype UTM (after the modification from
the original AM device), we point out that it should still be of
the same order of magnitude.  Therefore, since the input voltages
used did not exceed 10~V, the following results are in the small
modulation depth limit as described in Section
\ref{2model}\ref{2phasor2}.

A heterodyne detection scheme formed part of the measurement
apparatus, where a shunted beam was frequency-shifted by
$\omega_{h}/2\pi = 80$~MHz with an acousto-optic modulator (AOM).
If this heterodyne oscillator beam is represented by
$E_{\mathrm{HET}} = \gamma E_{\mathrm{IN}} e^{i \omega_{h}t}$ with
$\gamma \ll 1$ and is interfered with the modulated beam from
\eq{eqnelecfield}, then the detected power is given by:
\begin{eqnarray}
  \label{eqndetpower}
  P_{\mathrm{DET}} &=& E_{\mathrm{IN}}^2 \left[ P_{\mathrm{DC}} + P_{\omega_m} + P_{\omega_{h}} +
    P_{\omega_{h} -\omega_{m}} + P_{\omega_{h} +\omega_{m}}\right]; \nonumber \\
  P_{\mathrm{DC}} &=& \mathrm{cos}^2(\frac{\sigma}{2}); \nonumber \\
  P_{\omega_m} &=& 2 \mathrm{cos}(\frac{\sigma}{2})
    \Re\{\tilde{A}e^{i\omega_mt}\}; \nonumber \\
  P_{\omega_h} &=& 2 \mathrm{cos}(\frac{\sigma}{2})
    \Re\{\gamma e^{i\omega_ht}\}; \nonumber \\
  P_{\omega_{h}-\omega_{m}} &=& 2
    \Re \left[ \gamma (\tilde{A} + i\tilde{P})^\ast
    e^{i(\omega_h - \omega_m)t}\right] ;\nonumber \\
  P_{\omega_{h}+\omega_{m}} &=& 2
    \Re \left[ \gamma (\tilde{A} - i\tilde{P})
    e^{i(\omega_h + \omega_m)t} \right]
\end{eqnarray}
where the power components are split up according to their
respective frequencies.  $P_{\omega_m}$ represents the measurable
beat due to the presence of AM, $P_{\omega_h}$ is the beat between
the carrier and the heterodyne oscillator, and
$P_{\omega_{h}-\omega_{m}}$ and $P_{\omega_{h}+\omega_{m}}$
represent heterodyned copies of the two modulation sidebands. A
spectrum analyser was used to monitor the strength of these
frequency lines.  The use of the spectrum analyser facilitated the
assessment of the \emph{purity} of states produced by the UTM
prototype, by observing operating points that involved suppression
of one of these frequency lines.

An alternative set of measurables was obtained by electronically
mixing the heterodyne frequencies down to baseband via a
\emph{double-demodulation} scheme.  This gave a DC readout of the
PM and AM amplitudes at a selected beat phase.  A complete
description of the modulation strengths and quadratures was
obtained in this way, and was hence useful to characterise the
\emph{variability} of the UTM prototype.

The method of signal extraction via double demodulation is best
understood by reworking the last two components of
\eq{eqndetpower} to:
\begin{eqnarray}
  \label{eqndemodquad}
  P_{\omega_{h} - \omega_{m}} + P_{\omega_{h} + \omega_{m}}
  = 2 \mathrm{cos}(\frac{\sigma}{2})
    \left[ \Re\{\tilde{P} e^{i\omega_mt}\}
    \Im\{\gamma e^{i\omega_h t}\} \right. \nonumber \\
    \left. + \Re\{\tilde{A}e^{i\omega_mt}\}
    \Re\{\gamma e^{i\omega_h t}\} \right]
\end{eqnarray}
This signal is mixed down to base band by demodulating at
$\omega_{h}/2\pi = 80$~MHz and then $\omega_{m}/2\pi = 5$~MHz in
series.  The output of a demodulation is sensitive to the relative
phases of the signals being mixed: if they are exactly in-phase or
out-of-phase, the output will be maximally positive or negative;
if they are in quadrature, the output will be zero.
\eq{eqndemodquad} shows that the $\omega_{h}$ oscillator
components of the AM and PM portions are orthogonal (one is the
real component, the other is imaginary) so that the appropriate
choice of electrical oscillator phase can force the readout of PM
only, or AM only, or some linear combination of the two.
Similarly, the demodulation quadrature of the $\omega_{m}$ stage
determines the beat phase that the output signal is projected
onto.

The output DC component of the \emph{first} stage of demodulation
(extracted with a bias-T component) was used as an error signal
for locking the optical recombination phase of the heterodyne.
This error signal varies sinusoidally with respect to the optical
recombination phase, with a zero crossing where the optical
heterodyne beat ($P_{\omega_h}$) and the electronic demodulation
oscillator are in quadrature.  In other words, the feedback loop
"locks out" the heterodyne beat.  Comparing \eq{eqndetpower} and
\eq{eqndemodquad}, we see that the AM term has the same phase as
the $P_{\omega_h}$ term (that is, they both contain the terms
$\Re\{\gamma e^{i\omega_h t}\}$), so that the feedback loop also
locks out the AM component of modulation, leaving only the PM
component (which has a $\Im\{\gamma e^{i\omega_h t}\}$ term). This
is an important part of the process as, without a locking loop,
the demodulation phase of the circuit would be uncharacterised,
and would also be free to drift.

In the experiment, two separate double demodulation schemes were
used in order to measure PM and AM simultaneously; the
$\omega_{h}$ signal was split with a $90^{\mathrm{o}}$ electronic
splitter to ensure that the two double demodulation circuits
scanned orthogonal modulations.  Hence, when the heterodyne
locking loop was connected using a signal from one
double-demodulation circuit, this ensured that that particular
circuit was sensitive to PM and that the other was sensitive to
AM.

Regarding the second stage of demodulation, electronically phase
locking the signal generators was sufficient to time-stabilise the
demodulation quadrature, and a known modulation was used for
calibration.


\subsection{Results}
\label{3results3}

  \begin{figure}[!t]
   \centerline{\includegraphics{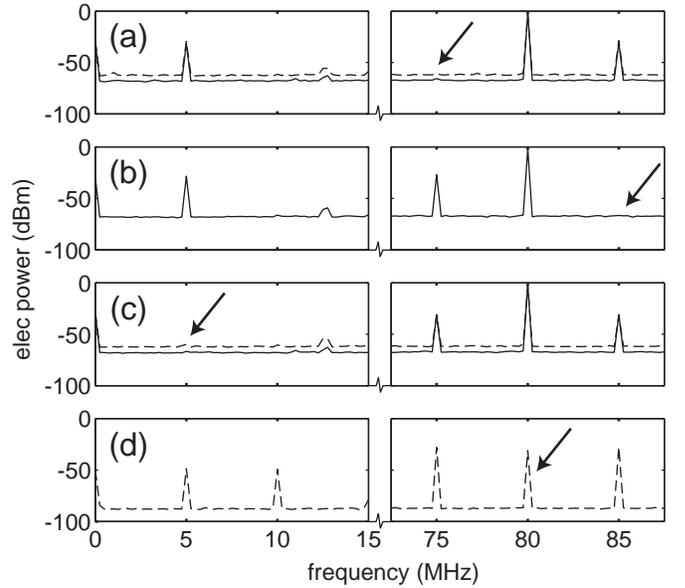}}
  \caption{\footnotesize{
        Spectrum analyser traces of heterodyne measurements of
        four "pure" operating points.  Frequencies of interest are
        5~MHz (direct AM beat), 80~MHz (heterodyne-carrier beat),
        and 75~MHz and 85~MHz (heterodyne-sideband beats).
        A solid line indicates data
        averaged over a few seconds, a dotted line indicates
        max-hold data acquired over approximately an hour, and
        arrows indicate lines that have been suppressed.  The
        plots are: (a) upper single-sideband (lower sideband
        suppressed by 34.9~dB); (b) lower single-sideband (upper sideband
        suppressed by 39.5~dB); (c) pure PM (AM beat suppressed by around 38~dB);
        (d) carrier suppression (heterodyne beat suppressed by around
        30~dB).}}
    \label{figoppoints}
  \end{figure}

\begin{figure}[!t]
   \centerline{\includegraphics{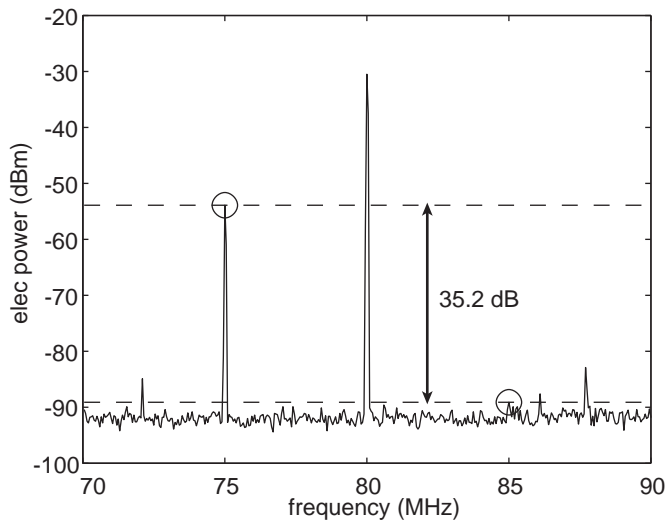}}
  \caption{\footnotesize{
        A more detailed spectrum analyser trace of sideband
        suppression, with 35.2~dB difference between
        sidebands.  Frequency lines at 72.1~MHz, 86.1~MHz and
        87.7~MHz are from radio interference with the electronic
        equipment.
  }}
    \label{figbestssb}
  \end{figure}

\fig{figoppoints} shows spectrum analyser traces for four
operating points, to demonstrate the suppression capabilities of
the device.  The suppression factors are 34.9~dB and 39.5~dB for
the left- and right-hand sidebands (\fig{figoppoints}(a) and
\fig{figoppoints}(b)).  The figures shown suppress the frequency
lines down to the electronic noise floor, so that higher
suppression factors may be possible.  A more detailed trace
exhibiting sideband suppression is given in \fig{figbestssb}, with
35.2~dB relative suppression recorded.

We measured around 38~dB suppression of the AM beat compared with
a similar-input-power pure-AM state (\fig{figoppoints}(c)).  In
this and the other diagrams in \fig{figoppoints}, we have shown
"max-hold" data, demonstrating that the device is highly stable,
maintaining these operating points without significant drift on
the timescale of hours.

In \fig{figoppoints}(d), the carrier is suppressed by selecting
the AM operating point and setting $\sigma = \pi$ (horizontal
polarisation).  The heterodyne measurement of the carrier is down
by around 30~dB, from which we can infer that the carrier power
(which goes as the square of the heterodyne measurement) is down
by 60~dB. This heterodyne measurement is further supported by
observing that the first and second harmonics of direct AM beat
are approximately equal, which is consistent with the fact that
the carrier (as measured by the heterodyne) is approximately 6~dB
weaker than the sidebands.

We were especially careful to validate this heterodyne measurement
of the carrier, because the directly detected power only dropped
by 40~dB.  The reason for this is that the majority of the
residual carrier power was now in a higher-order, odd spatial mode
(easily verified by looking at the intensity profile of the beam),
which did not interfere efficiently with the heterodyne oscillator
beam.  The higher-order modes were thought to result from a
spatially non-uniform polarisation of the light exiting the UTM
which, when subsequently passed through a polarising beam
splitter, produced modes reflecting the symmetry of the modulator.
As such, for applications where spatial mode interference is
important, the employment of a mode-cleaner cavity with free
spectral range equal to the modulation frequency (or one of its
integer divisors) should solve the problem.

\begin{figure}[!t]
  \centerline{\includegraphics{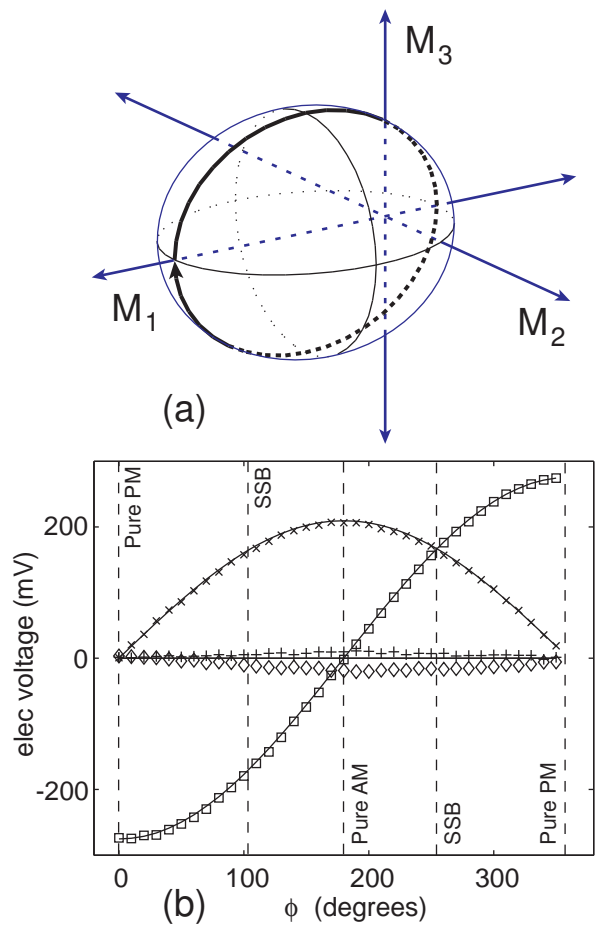}}
  \caption{\footnotesize{(a) Modulation ellipse showing trajectory of
  measurement sweep through parameter space, where the phase
  difference between electrical signals, $\phi = \phi_1 - \phi_2$, is varied through
  $360^{\circ}$.  (The ellipse has an (exaggerated) $\sigma$ of $\approx 53^{\circ}$.)
   (b) Double-demodulation measurements of PM (squares and
  diamonds for in-phase and quadrature components
  respectively) and AM (pluses and crosses for in-phase and
  quadrature components respectively), with corresponding
  theoretical predictions (solid lines).  The majority of modulation present is either
  PM or quadrature AM as predicted.  This plot corresponds to
  $\sigma \approx 75^{\circ}$; a value derived by measuring, from corresponding
  spectrum analyser data, the phase $\phi$ at which the
  single sideband operating points occur.  Using a result from Table
  I, the phase $\phi$ between the pure AM operating point and either of the
  single sideband operating points, is precisely equal to $\sigma$.
  }}
  \label{figsweep1}
\end{figure}

\begin{figure}[!t]
  \centerline{\includegraphics{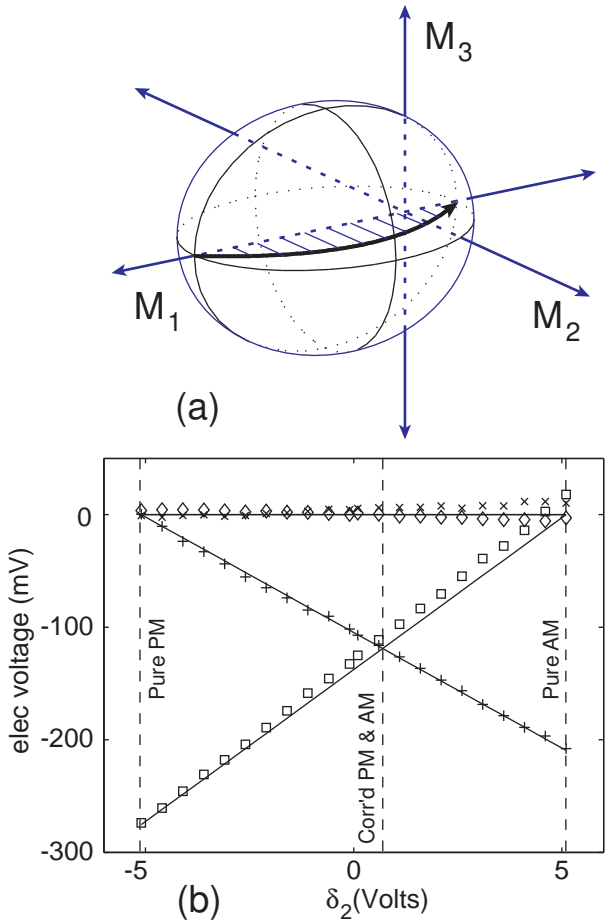}}
  \caption{\footnotesize{(a) Modulation ellipse showing trajectory of
  measurement sweep through parameter space, where the strength of
  one electrical signal, $\delta_2$, is varied from $+5.1$~V to
  $-5.1$~V. The sweep involves varying the overall input
  electrical power, so the modulation trajectory does not stay
  on the ellipse surface, and instead traces out a parabolic curve.
  (The ellipse has an (exaggerated) $\sigma$ of $\approx 53^{\circ}$.)
  (b) Double-demodulation measurements of PM (squares and
  diamonds for in-phase and quadrature components
  respectively) and AM (pluses and crosses for in-phase and
  quadrature components respectively), with corresponding
  theoretical predictions (solid lines).  There is diverging
  agreement in the PM which may be caused by imperfectly matched
  electrical impedances for the two crystals.  As in
  \fig{figsweep1}, the location of the correlated PM \& AM point
  is shifted toward the pure AM point by about $0.7$~V out of
  $5.1$~V and, referring to Table \ref{taboppoints}, this corresponds to
  a value of $\sigma \approx 75^{\circ}$.}}
  \label{figsweep2}
\end{figure}

In an attempt to confirm the variability of the UTM prototype, we
endeavoured to map significant paths in the modulation parameter
space.  \fig{figsweep1} and \fig{figsweep2} show the result of
sweeping through M-space in two cardinal directions: varying the
relative phase between the two (constant amplitude) electrical
signals; and varying the amplitude of one electrical signal while
keeping the other signal amplitude, and the relative phase,
constant.  For matters of calibration, the UTM was initially set
to the PM operating point (this initial state was used as the
reference in labelling "in-phase" and "quadrature" components),
and the relative gains and phases of the two double demodulation
circuits were measured and factored out via correspondence with
spectrum analyser data. The input polarisation was set to be
elliptical with a vertical component slightly larger than the
horizontal component.  The results show that changing the phase
between the two electrical signals produces AM in quadrature to
the original PM, whereas changing the amplitude of one electrical
signal generates AM in-phase with the original PM.  The data
points are generally in good agreement with the theoretical
predictions. Some small systematic errors are apparent, and are
thought to be associated with unmatched electrical and optical
impedances between the two crystals, and possibly to do with
slightly unmatched optical power levels probing each crystal (ie a
violation of our original assumption regarding polarisation
states).  Overall, the device is shown to be highly predictable,
and that one can "dial-up" a particular modulation state on call,
having calibrated the device initially.

It is instructive to notice that the cardinal operating points are
not evenly spaced when sweeping across the parameter space.  In
particular, we see that the two single-sideband operating points
are shifted toward the AM operating point in \fig{figsweep1}, and
the correlated PM \& AM operating point is similarly shifted in
\fig{figsweep2}, relative to the perpendicular. This is a product
of the non-circular polarisation of light input into the modulator
device, and can be understood by reviewing the modulation ellipses
in the two figures; the $M_2$ and $M_3$ axes intercept the ellipse
at points that are closer to the AM operating point than the PM.
In fact, this provides us with a means of calibration of the
relative transfer-function-amplitudes of the PM and AM
double-demodulation circuits: they are set by assuring that (in
\fig{figsweep1}) the PM and AM components are equal at the phase
$\phi$ where the single sideband is known to occur (information
which is obtained via comparison with spectrum analyser data).
Also, by using the identities in Table I, we can derive a value of
the polarisation parameter: $\sigma \approx 75^{\circ}$.


  \begin{figure}[!t]
    \centerline{\includegraphics{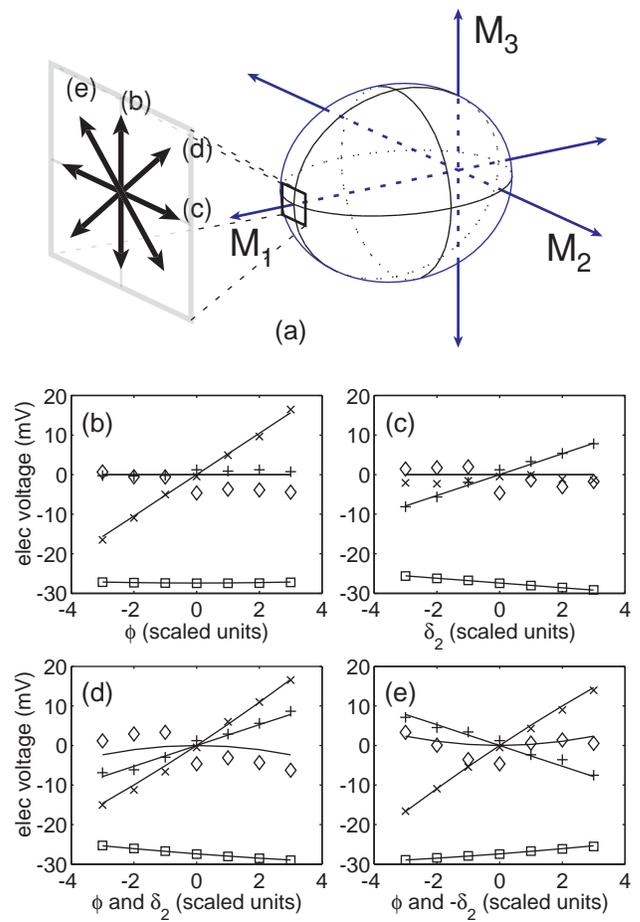}}
    \caption{\footnotesize{
        (a) Modulation ellipse showing short-range trajectories
        designed to map local region near PM operating point.  The
        electrical parameters $\phi = \phi_1 - \phi_2$ and
        $\delta_2$ were varied to achieve these results, with units
        worth $5^{\circ}$ and $0.44$~V respectively. (b), (c), (d)
        and (e) show double-demodulation measurements of PM (squares and
        diamonds for in-phase and quadrature components
        respectively) and AM (pluses and crosses for in-phase and
        quadrature components respectively), with corresponding
        theoretical predictions (solid lines).  Parameters varied were (b)
        $\phi$, (c) $\delta_2$, (d) $\phi$ and $\delta_2$
        together with the same polarity, and (e) $\phi$ and $\delta_2$
        together with opposite polarities.   \emph{Note:} the in-phase PM data
        points (squares) have been scaled down by a factor of 10 to fit in the
        diagram.  Also, the systematic error in the quadrature PM data is probably
        due to an electronic phase drift between signal
        generators, causing a small amount of in-phase PM data
        (which is an order of magnitude stronger) to
        couple across.}}
    \label{figoffsetlock}
  \end{figure}

\fig{figoffsetlock} shows scans of the local region near the PM
operating point, with deviations in four directions.  These data
were taken with an offset locking application in mind, as
discussed earlier. The in-phase PM component (in-phase by
definition) is by far the dominant signal, and has been scaled
down by a factor of 10 to fit in the graphs. As expected, the UTM
device can produce AM that is in-phase with the present PM, or AM
that is in quadrature with the present PM, by changing the
relative electrical signal amplitudes or phases respectively. In
addition, both of these parameters can be varied together or
oppositely, to give independent control over the two quadratures
of AM resulting. We note that in the case of producing both
in-phase and quadrature AM, a small component of quadrature PM
appears (in other words, the phase of the PM changes, relative to
that of the pure PM point used for the initial calibration). In
actual fact, the quadrature PM data has a noticeably larger
systematic error than the other three signals. This is most likely
due to pollution from its in-phase counterpart, caused by the
demodulation oscillator's phase drifting marginally (this was
observed to happen even in spite of the electronic phase locking
between signal generators).

A number of experimental difficulties deserve mentioning.  As
described in the context of the carrier suppression results, the
UTM produced a spatially varying polarisation state, which
produced a small percentage of higher-order spatial modes upon
selecting out a polarisation component (of order $1\%$ of the
overall power). This interfered with our ability to directly
measure the polarisation, which we did by measuring the overall
detected power while rotating a diagnostic half-wave plate placed
before the PBS. In this way, we measured a value of $\sigma
\approx 70^{\circ}$ for the results shown in Figs. \ref{figsweep1}
to \ref{figoffsetlock}, as compared to $\sigma \approx 75^{\circ}$
measured by inference from spectrum analyser data.  We assign a
reasonably large error to the direct measurement value due to the
presence of the higher-order spatial modes, and we favour
measuring $\sigma$ by inference from the data.

The overall optical path lengths of the modulating crystals were
found to change significantly as they warmed up after starting the
laser.  The dual crystal design of the device goes a long way
toward minimising this problem, and we have seen that the device
is stable once it has warmed up.  However, long term drift of the
crystal lengths is possible, having a direct effect on the
polarisation state leaving the device.  We propose that, in
circumstances where this becomes a problem, a feedback loop be
employed to lock the polarisation state.  A possible scheme would
see the power level out of one PBS port monitored and used as a
feedback signal (minus an offset equal to the desired power level)
to the DC optical path length of one of the crystals, thus
compensating for a mechanical path length change by feeding back
to the refractive index.

Finally, the electrical impedances of the two crystals were not
well matched for two reasons: the electronics were not identical;
and the crystals themselves had "good" and "bad" spots which
generated varying levels of modulation.  In general terms, careful
alignment can largely overcome this problem and return the two
crystals to an equal footing.  The differing electrical impedances
complicate the issue of generating a local oscillator with the
phase-tracking property discussed earlier, since the voltages
$V_{1,2}$ and modulation depths $\delta_{1,2}$ are not then
related by the same factor.  We submit that the device tested here
is merely a prototype, and that careful management will suffice to
deal with these issues as the need arises.


\section{Conclusion}
\label{4conc}

We presented a thorough investigation, both theoretical and
experimental, of a prototype Universally Tunable Modulator (UTM).
The electrical-to-optical transfer function of the device was
derived, both in terms of electrical and optical phasor notation,
and using a geometric phase "Modulation Sphere" representation.
Both pictures were shown to have merit, and a set of cardinal
modulations were described in each notation.  We reported on an
experiment to characterise the prototype UTM, which involved using
dual double-demodulation circuits to measure both AM and PM
components simultaneously.  Data sets were obtained and analysed
to illustrate the variability and purity characteristics of the
device; the device was shown to be highly predictable and capable
of highly pure states.  Applications for the UTM were discussed.

\section{Acknowledgements}
The authors are grateful to Russell Koehne for his skill and
labour in modifying the original AM device. Stan Whitcomb thanks
the gravitational wave research group at the Australian National
University for their support and hospitality during his stay at
the ANU.

This material is based on work supported by the Australian
Research Council, and supported in part by the United States
National Science Foundation under Cooperative Agreement No.
PHY-0107417.  This paper has been assigned LIGO Laboratory
document number LIGO-P030031-00R.

\appendix

\section{Appendices}
\subsection{Derivation of UTM transfer function}\label{secapp1}

This is an outline of the derivation of \eq{eqntf}.  First, we
find the (more general) transfer function of the UTM when we allow
the input beam to have any polarisation.  At the end, we will
simplify to the subcase described in the main text.

The input beam's polarisation will be characterised by two
electric field phasor components, $\tilde{L} = Le^{i\sigma_L}$ and
$\tilde{R} = Re^{i\sigma_R}$, as defined in a set of left-diagonal
and right-diagonal spatial coordinate axes (with unit vectors
$\hat{L}$ and $\hat{R}$ respectively).  The electric field exiting
the UTM can be written as a vector (to include polarisation
information) as:
\begin{equation}\label{eqnapp1}
    \vec{E}_{\mathrm{Exiting UTM}} =
    \left[\tilde{L}e^{i\Re\{\tilde{\delta}_1e^{i\omega_mt}\}}\hat{L}
    + \tilde{R}e^{i\Re\{\tilde{\delta}_2e^{i\omega_mt}\}}\hat{R}\right]
    e^{i\omega t}.
\end{equation}

(From hereon in, we will suppress the $e^{i\omega t}$ term for the
sake of brevity.)  Upon passing through the vertically aligned
linear polariser (equivalent to taking a dot product with the
vector $(\hat{L}+\hat{R})/\sqrt{2}$), and expanding the complex
exponentials to first order (hence assuming
$|\tilde{\delta}_{1,2}| \ll 1$), the vertical electric field
amplitude becomes:
\begin{equation}\label{eqnapp2}
    E_{\mathrm{OUT}} =
    \frac{1}{\sqrt{2}}\left[\tilde{L}(1+i\Re\{\tilde{\delta}_1e^{i\omega_mt}\})
    + \tilde{R}(1+i\Re\{\tilde{\delta}_2e^{i\omega_mt}\}\right].
\end{equation}

Next, we collect DC terms and oscillating terms, and factor out
the overall optical phase:
\begin{eqnarray}\label{eqnapp3}
     E_{\mathrm{OUT}} &=&
    \frac{\tilde{L}+\tilde{R}}{\sqrt{2}}\left[ 1 +
    \frac{i(|L^2+\tilde{L}\tilde{R}^*)}{|\tilde{L}+\tilde{R}|^2}
    \Re \{\tilde{\delta}_1 e^{i\omega_mt} \} \right. \nonumber \\
     &+& \left.
    \frac{i(\tilde{R}\tilde{L}^*+R^2)}{|\tilde{L}+\tilde{R}|^2}
    \Re \{\tilde{\delta}_2 e^{i\omega_mt} \}
    \right].
\end{eqnarray}

Now, we work to separate real and imaginary components for the
oscillating terms:
\begin{eqnarray}\label{eqnapp4}
     E_{\mathrm{OUT}} &=&
    \frac{\tilde{L}+\tilde{R}}{\sqrt{2}}\left[ 1 +
    \Re \left( \frac{\Im\{\tilde{R}\tilde{L}^*\}(\tilde{\delta}_1-\tilde{\delta}_2)}
    {|\tilde{L}+\tilde{R}|^2}e^{i\omega_mt}
    \right) \right. \nonumber \\
    &+& \left. i \Re \left( \frac{\Re
    \{\tilde{R}\tilde{L}^*\}(\tilde{\delta}_1+\tilde{\delta}_2)+
    L^2\tilde{\delta}_1+R^2\tilde{\delta}_2)}
    {|\tilde{L}+\tilde{R}|^2} e^{i\omega_mt}
    \right) \right].
\end{eqnarray}

These real and imaginary oscillating terms correspond to AM and PM
respectively, so we parameterize via:
\begin{equation}\label{eqnapp5}
    E_{\mathrm{OUT}} = \frac{|\tilde{L}+\tilde{R}|}{\sqrt{2}} +
    \Re\{\tilde{A}e^{i\omega_mt}\} + i
    \Re\{\tilde{P}e^{i\omega_mt}\}
\end{equation}
with
\begin{eqnarray}\label{eqnapp6}
    \tilde{P} &=& \Re \left( \frac{\Re
    \{\tilde{R}\tilde{L}^*\}(\tilde{\delta}_1+\tilde{\delta}_2)+
    L^2\tilde{\delta}_1+R^2\tilde{\delta}_2)}
    {\sqrt{2}|\tilde{L}+\tilde{R}|} \right)   \nonumber \\
    \tilde{A} &=& \Re \left( \frac{\Im\{\tilde{R}\tilde{L}^*\}(\tilde{\delta}_1-\tilde{\delta}_2)}
    {\sqrt{2}|\tilde{L}+\tilde{R}|}\right)
\end{eqnarray}
where the net optical phase shift has been discarded.  Eqs.
\ref{eqnapp5} and \ref{eqnapp6} constitute the general UTM
transfer function for arbitrary input polarisation.  Note that the
PM component does not have the property of tracking the phase of
the sum of the input electrical signals, since it generally
depends on these inputs in different proportions.  This is one of
the reasons why we chose to restrict the polarisation to a subset.

If we choose said subset, $L=R=E_{\mathrm{IN}}/\sqrt{2}$ (equal
power in the two polarisation axes), and define $\sigma =
\sigma_R-\sigma_L$ as the phase between the two polarisation
components, then the equations reduce to:
\begin{equation}\label{eqnapp7}
    E_{\mathrm{OUT}} =  E_{\mathrm{IN}} \mathrm{cos}
    \left(\frac{\sigma}{2}\right) +
    \Re\{\tilde{A}e^{i\omega_mt}\} + i
    \Re\{\tilde{P}e^{i\omega_mt}\}
\end{equation}
and
\begin{eqnarray}\label{eqnapp8}
    \tilde{P} &=& \frac{E_{\mathrm{IN}}}{2} \mathrm{cos}
    \left(\frac{\sigma}{2}\right)\left(\tilde{\delta}_1+\tilde{\delta}_2\right)
    \nonumber \\
    \tilde{A} &=& \frac{E_{\mathrm{IN}}}{2} \mathrm{sin}
    \left(\frac{\sigma}{2}\right)\left(\tilde{\delta}_1-\tilde{\delta}_2\right).
\end{eqnarray}

Eqs. \ref{eqnelecfield} and \ref{eqntf} in the text are the same
as these, but with a dimensionless definition of $\tilde{P}$ and
$\tilde{A}$.

\subsection{Derivation of UTM transfer function for Q-space and M-space parameters}
\label{secapp2}

This is an outline of the derivation of \eq{eqntfellipse}.  As
with Appendix \ref{secapp1}, we derive the Q-space to M-space
transfer function of the UTM for any input light polarisation
state. At the end, we reduce the equations to the case where the
allowed polarisations are restricted.  The derivation consists of
transforming parameters $\tilde{P}$ and $\tilde{A}$ to M-space
parameters $(M_1,M_2,M_3)$ using \eq{eqngeospherem}, and
transforming parameters $\tilde{\delta}_1$ and $\tilde{\delta}_2$
to Q-space parameters $(Q_1,Q_2,Q_3)$ using \eq{eqngeospherep},
and hence converting \eq{eqnapp6} from one set of coordinates to
another.

Firstly, it is convenient to write down a few intermediate terms
using \eq{eqnapp6} as a starting point.  We also convert to
Q-space parameters (via \eq{eqngeospherep}) as we go.

\begin{eqnarray}\label{eqnapp9}
    P^2 = \tilde{P}^*\tilde{P} = \frac{\Re
    \{\tilde{R}\tilde{L}^*\}^2 [Q_0 + Q_2] + 2L^2R^2Q_2}{2|\tilde{L}+\tilde{R}|^2}
    \nonumber \\
    + \frac{\frac{1}{2}[(L^4 + R^4)Q_0 + (L^4 - R^4)Q_1]}{2|\tilde{L}+\tilde{R}|^2}
    \nonumber \\
    + \frac{\Re \{\tilde{R}\tilde{L}^*\}[(L^2+R^2)Q_0 +(L^2-R^2)Q_1 +
    (L^2+R^2)Q_2]} {2|\tilde{L}+\tilde{R}|^2}
\end{eqnarray}
\begin{equation}\label{eqnapp10}
    A^2 = \tilde{A}^*\tilde{A} = \frac{\Im
    \{\tilde{R}\tilde{L}^*\}^2 [Q_0 - Q_2]}
    {2|\tilde{L}+\tilde{R}|^2}
\end{equation}
\begin{eqnarray}\label{eqnapp11}
    \Re\{\tilde{P}\tilde{A}^*\} = \frac{\Re\{\tilde{R}\tilde{L}^*\}
    \Im\{\tilde{R}\tilde{L}^*\}Q_1}{2|\tilde{L}+\tilde{R}|^2}
    \nonumber \\
    + \frac{\frac{1}{2}\Im\{\tilde{R}\tilde{L}^*\}\left[(L^2-R^2)Q_0 +(L^2+R^2)Q_1
    -(L^2-R^2)Q_2\right]}{2|\tilde{L}+\tilde{R}|^2}
\end{eqnarray}
\begin{eqnarray}\label{eqnapp12}
    \Im\{\tilde{P}\tilde{A}^*\} =
    \frac{-\Im\{\tilde{R}\tilde{L}^*\}\left[\Re\{\tilde{R}\tilde{L}^*\}
    + \frac{1}{2}(L^2+R^2)\right]Q_3
    }{2|\tilde{L}+\tilde{R}|^2}
\end{eqnarray}

Things can be simplified somewhat by using a Stokes' parameter
representation for the polarisation from this point. Here, we
(unusually) define the Stokes' parameters in terms of
left-diagonal and right-diagonal electric field components
(equivalent to rotating the usual X-Y axes anti-clockwise by
$45^{\circ})$:
\begin{eqnarray} \label{eqnstokes}
    S_0 &=& R^2 + L^2 \nonumber \\
    S_1 &=& R^2 - L^2 \nonumber \\
    S_2 &=& 2RL\mathrm{cos}(\sigma) = 2\Re \{ \tilde{R} \tilde{L}^* \} \nonumber \\
    S_3 &=& 2RL\mathrm{sin}(\sigma) = 2\Im \{ \tilde{R} \tilde{L}^* \}
\end{eqnarray}
when the previous four equations become:
\begin{eqnarray} \label{eqnapp14}
    P^2 &=& \frac{1}{4} \left[S_0 Q_0 - S_1 Q_1 + S_2 Q_2 \right. \nonumber \\
    &-& \left. \frac{S_3^2}{2(S_0 + S_2)}(Q_0 - Q_2)\right] \nonumber \\
    A^2 &=& \frac{1}{8} \frac{S_3^2}{S_0+S_2}(Q_0-Q_2) \nonumber \\
    \Re\{\tilde{P}\tilde{A}^*\} &=& \frac{1}{8}\left[ S_3 Q_1 -
    \frac{S_3 S_1}{S_0 + S_2}(Q_0 - Q_2) \right] \nonumber \\
    \Im\{\tilde{P}\tilde{A}^*\} &=& -\frac{1}{8} S_3 Q_3
\end{eqnarray}

These intermediate quantities are then converted to M-space
parameters (via \eq{eqngeospherem}).  We write the result here in
matrix form:
\begin{equation}\label{eqnapp15}
    \begin{bmatrix} M_0 \\ M_1 \\ M_2 \\ M_3 \end{bmatrix} =
    \frac{1}{4}
    \begin{bmatrix}
        S_0 & -S_1 & S_2 & 0 \\
        S_2 + \frac{S_1^2}{S_0 + S_2} & -S_1 & S_0 - \frac{S_1^2}{S_0 +
            S_2}& 0 \\
        \frac{-S_1 S_3}{S_0 + S_2} & S_3 & \frac{S_1 S_3}{S_0 +
            S_2}& 0 \\
        0 & 0 & 0 & -S_3
    \end{bmatrix}
    \begin{bmatrix} Q_0 \\ Q_1 \\ Q_2 \\ Q_3 \end{bmatrix}
\end{equation}

\eq{eqnapp15} is the general Q-space to M-space transfer function
for the UTM with any input polarisation.  It gives further insight
into the manner in which choosing $L\ne R$ affects the properties
of the system.  The parameters $M_1$ and $M_2$ depend on both
$Q_1$ and $Q_2$, which means that a sphere in Q-space will be
distorted in the $M_1$-$M_2$ plane by the transformation.  The
resulting surface is an ellipsoid whose major axis is no longer
collinear with the $M_1$-axis, but is at an angle subtended in the
$M_1$-$M_2$ plane.  This distortion certainly interferes with most
of the favourable properties of the system.  For example, none of
the poles of the ellipse coincide with a coordinate axis, making
any of the six operating points discussed more complicated to
find.

The restriction on polarisation states discussed in the text is
equivalent to setting $S_1$ to zero (which in turn forces
$S_0=E_{\mathrm{IN}}^2$,
$S_2=E_{\mathrm{IN}}^2\mathrm{cos}(\sigma)$ and
$S_3=E_{\mathrm{IN}}^2\mathrm{sin}(\sigma)$), when \eq{eqnapp15}
reduces to:
\begin{equation}\label{eqnapp16}
    \begin{bmatrix} M_0 \\ M_1 \\ M_2 \\ M_3 \end{bmatrix} =
    \frac{E_{\mathrm{IN}}^2}{4}
    \begin{bmatrix}
        1 & 0 & \mathrm{cos}(\sigma) & 0 \\
        \mathrm{cos}(\sigma) & 0 & 1 & 0 \\
        0 & \mathrm{sin}(\sigma) & 0 & 0 \\
        0 & 0 & 0 & -\mathrm{sin}(\sigma)
    \end{bmatrix}
    \begin{bmatrix} Q_0 \\ Q_1 \\ Q_2 \\ Q_3 \end{bmatrix}
\end{equation}

Once again, \eq{eqntfellipse} from the text is the dimensionless
equivalent of this, where $E_{\mathrm{IN}}$ is external to the
definition of the M-space parameters.






\newpage





\end{document}